\documentclass[12pt]{iopart}
\usepackage{iopams}
\usepackage{graphicx}
\newcommand{\veps}{\varepsilon}
\newcommand{\rhm}{\rho_{\mathrm{m}}}
\newcommand{\prm}{p_{\mathrm{m}}}
\newcommand{\dph}{\dot{\phi}}
\newcommand{\pat}{\partial}
\newcommand{\rhmz}{\rho_{\mathrm{m}0}}
\newcommand{\prmz}{p_{\mathrm{m}0}}
\begin{document}
\title{Regular accelerating Universe without dark energy}

\author{A.V. Minkevich$^{1,2}$, A.S. Garkun$^1$ and V.I. Kudin$^3$}

\address{$^1$Department of Theoretical Physics, Belarussian State University, Belarus}
\address{$^2$Department of Physics and Computer Methods, Warmia and Mazury University in Olsztyn,
Poland}
 \address{$^3$Department of Technical Physics, Belarussian National Technic University, Belarus}
 \eads{\mailto{minkav@bsu.by}, \mailto{garkun@bsu.by}}

\begin{abstract}
Homogeneous isotropic cosmological models with two torsion functions filled with scalar fields and
usual gravitating matter are built and investigated in the framework of the Poincar\'e gauge theory
of gravity. It is shown that by certain restrictions on indefinite parameters of gravitational
Lagrangian the cosmological equations at asymptotics contain an effective cosmological constant
that can explain observable acceleration of cosmological expansion. The behavior of inflationary
cosmological solutions at extremely high energy densities is analyzed, regular bouncing solutions
are obtained. The role of the space-time torsion provoking the acceleration of cosmological
expansion is discussed.
\end{abstract}
\pacs{04.50.+h; 98.80.Cq; 11.15.-q; 95.36.+x}
 \submitto{\CQG}

\section{Introduction}

The discovery of the acceleration of cosmological expansion at present epoch is the most principal
achievement of observational cosmology at last time [1]. By using Friedmann cosmological equations
of General Relativity theory (GR) in order to explain accelerating cosmological expansion, the
notion of dark energy (or quintessence) was introduced in cosmology. According to obtained
estimations, approximately 70\% of energy in our Universe is related with some hypothetical form of
gravitating matter with negative pressure --- ``dark energy'' --- of unknown nature. Previously a
number of investigations devoted to dark energy problem were carried out (see review [2]).
According to widely known opinion, the dark energy is associated with cosmological term. If the
cosmological term is related to  the vacuum energy density, it is necessary to explain, why it has
the value close to critical energy density at present epoch (see for example [3]).

The present paper is devoted to investigation of the ``dark energy'' problem in the framework of
the Poincar\'e gauge theory of gravity (PGTG), which is a natural generalization of Einsteinian GR
by applying  the gauge approach to the theory of gravitational interaction (see for example [11]).
In fact the generalization of GR leading to PGTG is necessary, if one supposes that the Lorentz
gauge field corresponding to fundamental group in physics -- the Lorentz group -- exists in the
nature (see [4]). According to PGTG the physical space-time possesses the structure of
Riemann-Cartan continuum with curvature and torsion. The investigation of isotropic cosmology built
in the framework of PGTG (see [4-6] and references herein) shows that the gravitational interaction
in PGTG, unlike GR and Newton's gravitation theory, can have the repulsion character in the case of
gravitating systems with positive energy densities and pressures satisfying energy dominance
condition. So, the gravitational repulsion effect takes place at extreme conditions (extremely high
energy densities and pressures) preventing the appearance of cosmological singularity in
homogeneous isotropic models (HIM) [5]. According to generalized cosmological Friedmann equations
(GCFE) for HIM deduced in the framework of PGTG, all cosmological solutions including inflationary
solutions are regular in metrics, Hubble parameter, its time derivative and have bouncing
character. Properties of discussed HIM in PGTG coincide practically with that of GR at sufficiently
small energy densities, which are much less in comparison with limiting (maximum) energy density
for such models. By including cosmological term of corresponding value to GCFE, we can obtain
regular cosmological solutions with observable accelerating expansion stage.  However, like GR, the
problem of dark energy in such theory is not solved.

From geometrical point of view, the structure of HIM in PGTG can be more complicated in comparison
with models describing by GCFE. In fact in the case of homogeneous isotropic models the torsion
tensor $S^\lambda{}_{\mu\nu}=-S^\lambda{}_{\nu\mu}$ can have the following non-vanishing components
[7, 8]: $S^1{}_{10}=S^2{}_{20}=S^3{}_{30}=S_{1}(t)$, $\displaystyle S_{123} = S_{231} = S_{312} =
S_{2}(t)\frac{R^3r^2}{\sqrt{1-kr^2}} \sin{\theta}$, where $S_{1}$ and $S_{2}$ are two torsion
functions of time, spatial spherical coordinates are used. The functions $S_{1}$ and $S_{2}$ have
different properties with respect to transformations of spatial inversions, namely, the function
$S_{2}(t)$ has pseudoscalar character. The GCFE follow from gravitational equations of PGTG for HIM
together with $S_{2}=0$. Obtained physical consequences of GCFE have principal character. However,
it is necessary to note that gravitational equations of PGTG for HIM have also other solution with
non-vanishing function $S_{2}$.

The HIM with two torsion functions filled with scalar fields and usual gravitating matter are
studied below in the frame of PGTG in connection with the dark energy problem. Following to [9], in
Section 2 cosmological equations for such HIM are introduced. In Section 3 the solutions
asymptotics of cosmological equations is analyzed. In Section 4 the bouncing character of
inflationary cosmological solutions is examined.

\section{Cosmological equations for HIM with two torsion functions in PGTG}

At first, let us mention some general relations of the PGTG. Gravitational field is described in
the frame of PGTG by means of the orthonormalized tetrad $h^i{}_\mu$ and anholonomic Lorentz
connection $A^{ik}{}_\mu$ (tetrad and holonomic indices are denoted by latin and greek
respectively); corresponding field strengths are torsion $S^i{}_{\mu\nu}$ and curvature
$F^{ik}{}_{\mu\nu}$ tensors defined as
\[
S^i{}_{\mu \,\nu }  = \partial _{[\nu } \,h^i{}_{\mu ]}  -
h_{k[\mu } A^{ik}{}_{\nu ]}\,,
\]
\[
F^{ik}{}_{\mu\nu }  = 2\partial _{[\mu } A^{ik}{}_{\nu ]}  +
2A^{il}{}_{[\mu } A^k{}_{|l\,|\nu ]}\,.
\]
We will consider the PGTG
based on the following general form of gravitational Lagrangian
\begin{eqnarray}\label{1}\fl
{\cal L}_{\rm G}= h\left[f_0\,
F+F^{\alpha\beta\mu\nu}\left(f_1\:F_{\alpha\beta\mu\nu}
                +f_2\: F_{\alpha\mu\beta\nu}+f_3\:F_{\mu\nu\alpha\beta}\right)
        + F^{\mu\nu}\left(f_4\:F_{\mu\nu}+f_5\: F_{\nu\mu}\right)+f_6\:F^2
    \right.
\nonumber \\ \left.
    +S^{\alpha\mu\nu}\left(a_1\:S_{\alpha\mu\nu}+a_2\: S_{\nu\mu\alpha}\right)
    +a_3\:S^\alpha{}_{\mu\alpha}S_\beta{}^{\mu\beta}\right],
\end{eqnarray}
where $h=\det{\left(h^i{}_\mu\right)}$,
$F_{\mu\nu}=F^{\alpha}{}_{\mu\alpha\nu}$, $F=F^\mu{}_\mu$, $f_i$
($i=1,2,\ldots,6$), $a_k$ ($k=1,2,3$) are indefinite parameters,
$f_0=(16\pi G)^{-1}$, $G$ is Newton's gravitational constant.
Gravitational equations of PGTG obtained from the action integral
$ I = \int {\left( {{\cal L}_g + {\cal L}_m } \right)\,} d^4 x$,
where  ${\cal L}_m$ is the Lagrangian of matter, contain the
system of 16+24 equations corresponding to gravitational variables
$h^i{}_\mu$ and $A^{ik}{}_\mu$.

Any homogeneous isotropic gravitating system in PGTG is characterized in general case by three
functions of time: the scale factor of Robertson-Walker metrics $R$ and two torsion functions
$S_{1}$ and $S_{2}$. Below the spherical coordinate system is used and the tetrad is taken in
diagonal form. Then the curvature tensor has the following non-vanishing tetrad components denoted
by means of the sign \^{} :
\begin{eqnarray}
&&
  F^{\hat 0\hat 1}{}_{\hat 0\hat 1}  = \,F^{\hat 0\hat 2}{}_{\hat 0\hat 2}
    = F^{\hat 0\hat 3}{}_{\hat 0\hat 3}
    \equiv A_1,
  \qquad
  F^{\hat 1\hat 2}{}_{\hat 1\hat 2}
    = F^{\hat 1\hat 3}{}_{\hat 1\hat 3}  = F^{\hat 2\hat 3}{}_{\hat 2\hat 3}
    \equiv A_2, 
\nonumber\\
&&
  F^{\hat 0\hat 1}{}_{\hat 2\hat 3}  = \,F^{\hat 0\hat 2}{}_{\hat 3\hat 1}
    = F^{\hat 0\hat 3}{}_{\hat 1\hat 2}
    \equiv A_3,
  \qquad
  F^{\hat 3\hat 2}{}_{\hat 0\hat 1} = F^{\hat 1\hat 3}{}_{\hat 0\hat 2}
    = F^{\hat 2\hat 1}{}_{\hat 0\hat 3} \equiv A_4 
\nonumber 
\end{eqnarray}
with
\begin{equation}\label{2}
\eqalign{
    A_1=\dot{H}+H^2-2HS_1-2\dot{S}_1,
    \\ 
    A_{2}  = \frac{k} {{R^2 }} + \left( {H - 2S_1 } \right)^2  - S_2^2,
    \\ 
    A_{3}  = 2\left( {H - 2S_1 } \right)S_2,
    \\ 
    A_{4}  = \dot S_2+HS_2,
}
\end{equation}
where $H=\dot{R}/R $ is the Hubble parameter and a dot denotes the
differentiation with respect to time.

The Bianchi identities in this case are reduced to two following
relations:
\begin{equation}\label{3}
\eqalign{
    \dot A_{2}  + 2H\left( {A_{2}  - A_{1} } \right) + 4S_1 A_{1}
        + 2S_2 A_{4}  = 0,
    \\ 
    \dot A_{3}  + 2H\left( {A_{3}  - A_{4} } \right) + 4S_1 A_{4}
        - 2S_2 A_{1}  = 0.
}
\end{equation}
The system of gravitational equations for HIM corresponding to gravitational Lagrangian (\ref{1})
has the following form
\begin{equation}\label{4}\fl
a\left( {H - S_1 } \right)S_1  - 2bS_2^2  - 2f_0 A_{2}  + 4f\left(
{A_{1}^2 - A_{2}^2 } \right) + 2q_2 \left( {A_{3}^2  - A_{4}^2 }
\right) =  - \frac{\rho } {3},
\end{equation}
\begin{equation}\label{5}\fl
a\left( {\dot S_1  + 2HS_1  - S_1^2 } \right) - 2bS_2^2  - 2f_0
\left( {2A_{1} + A_{2} } \right) - 4f\left( {A_{1}^2  - A_{2}^2 }
\right) - 2q_2 \left( {A_{3}^2  - A_{4}^2 } \right) = p,
\end{equation}
\begin{equation}\label{6}\fl
f\left[ {\dot A_{1}  + 2H\left( {A_{1}  - A_{2} } \right) + 4S_1
A_{2} } \right] + q_2 S_2 A_{3}  - q_1 S_2 A_{4}  + \left( {f_0  +
\frac{a} {8}} \right)S_1  = 0,
\end{equation}
\begin{equation}\label{7}\fl
q_2 \left[ {\dot A_{4}  + 2H\left( {A_{4}  - A_{3} } \right) +
4S_1 A_{3} } \right] - 4f\,S_2 A_{2} - 2q_1 S_2 A_{1}  - \left(
{f_0  - b} \right)S_2  = 0,
\end{equation}
where
\begin{eqnarray}
  a = 2a_1  + a_2  + 3a_3, \qquad b = a_2  - a_1,
\nonumber\\
  f = f_1  + \frac{{f_2 }} {2} + f_3  + f_4  + f_5  + 3f_{6}\, ,
\nonumber\\
  q_1  = f_2  - 2f_3  + f_4  + f_5  + 6f_{6}, \qquad q_2  = 2f_1  - f_2 ,
\nonumber
\end{eqnarray}
$\rho$ is the energy density, $p$ is the pressure and the average of spin distribution of
gravitating matter is supposed to be equal to zero. Equations (\ref{4})--(\ref{5}) lead to
generalization of Friedmann cosmological equations of GR, which does not contain high derivatives
for the scale factor $R$, if $a=0$ (see below). Moreover, equations (\ref{6})--(\ref{7}) take more
symmetric form, if $2f=q_1+q_2$. Then by using the Bianchi identities (\ref{3}), the system of
gravitational equations for HIM take the following form:
\begin{equation}\label{8}
 - 2b\,S_2^2  - 2f_0 A_{2}  + 4f\left( {A_{1}^2  - A_{2}^2 } \right) + 2q_2 \left( {A_{3}^2  - A_{4}^2 } \right)
 =  - \frac{1}{3}\rho,
\end{equation}
\begin{equation}\label{9}
 - 2b\,S_2^2  - 2f_0 \left( {2A_{1}  + A_{2} } \right) - 4f\left( {A_{1}^2  - A_{2}^2 } \right) - 2q_2
    \left( {A_{3}^2  - A_{4}^2 } \right) = p,
\end{equation}
\begin{equation}\label{10}
    f\left[ {\left( {\dot A_{1}  + \dot A_{2} } \right) + 4S_1 \left( {A_{1}  + A_{2} }
        \right)} \right] + q_2\, S_2 \left( {A_{3}  + A_{4} } \right) + f_0 S_1  = 0,
\end{equation}
\begin{equation}\label{11}
    q_2 \left[ {\left( {\dot A_{3}  + \dot A_{4} } \right) + 4S_1 \left( {A_{3}  + A_{4} }
        \right)} \right] - 4f\,S_2 \left( {A_{1}  + A_{2} } \right)
    - \left( {f_0  - b} \right)S_2  = 0.
\end{equation}
The system of equations (\ref{8})--(\ref{11}) together with
definition of curvature functions (\ref{2}) is the base of our
investigation of HIM below. Note also that the conservation law for
spinless matter has the usual form:
\begin{equation}\label{12}
    \dot \rho  + 3H\left( {\rho  + p} \right) = 0.
\end{equation}
In order to investigate inflationary cosmological models we will consider below HIM filled with
non-interacting scalar field $\phi$ minimally coupled with gravitation and gravitating matter with
equation of state in the form $p_m=p_m(\rho_m)$ (values of gravitating matter are denoted by means
of index ``${m}$''). Then the energy density $\rho$ and the pressure $p$ take the form
\begin{equation}
\label{13} \rho=\frac{1}{2}\dot{\phi}^2+V+\rho_m \quad (\rho>0),
\quad p=\frac{1}{2}\dot{\phi}^2-V+p_m,
\end{equation}
where $V=V(\phi)$ is a scalar field potential. By using the scalar
field equation in homogeneous isotropic space
\begin{equation}
\label{14} \ddot{\phi}+3H\dph=-\frac{\pat V}{\pat\phi}
\end{equation}
we obtain from  (\ref{12})--(\ref{13}) the conservation law for gravitating matter
\begin{equation}
\label{15}\dot{\rho}_m+3H\left(\rhm+\prm\right)=0.
\end{equation}
From (\ref{8})--(\ref{9}) follows that
\begin{equation}\label{16}
    A_{1}  + A_{2}  = \frac{1} {12f_0 }\left( \rho  - 3p\right) - \frac{b}{f_0}S_2^2.
\end{equation}
By using  (\ref{16}) and the formula following from definition of curvature functions $A_3$ and
$A_4$
\[
    A_{3}^2  - A_{4}^2 = 4A_{2}\, S_2^2  - 4\left( {\frac{k}{{R^2 }} - S_2^2 } \right)S_2^2
        - \left( {\dot S_2  + HS_2 } \right)^2 ,
\]
we find from gravitational equations (\ref{8})--(\ref{9}) the following expressions for $A_1$ and
$A_2$:
\begin{eqnarray}\label{17}\fl
    A_{1}  = -\frac{1} {{12f_0 Z}}
        \left[
            \rho  + 3p - \frac{\alpha } {2} \left( {\rho  - 3p - 12bS_2^2 } \right)^2
        \right]
\nonumber \\
        - \frac{\alpha \varepsilon }{Z}\left( {\rho  - 3p - 12bS_2^2 } \right)S_2^2
        + \frac{{3\alpha \varepsilon f_0 }} {Z}
            \left[ {\left( {HS_2  + \dot S_2 } \right)^2
                + 4\left( {\frac{k}{{R^2 }} - S_2^2 } \right)S_2^2 }
            \right],
\end{eqnarray}
\begin{eqnarray}\label{18}\fl
   A_{2} = \frac{1} {{6f_0 Z}}
        \left[
            {\rho  -6 b S_2^2
            + \frac{\alpha }{4} \left( {\rho  - 3p - 12bS_2^2 } \right)^2 }
        \right]
\nonumber\\
        - \frac{{3\alpha \varepsilon f_0 }} {Z}
            \left[
                {\left( {HS_2  + \dot S_2 } \right)^2
                + 4\left( {\frac{k}{{R^2 }} - S_2^2 } \right)S_2^2 }
            \right],
\end{eqnarray}
where $Z\equiv 1+\alpha\left( \rho - 3p - 12\left( {b + \varepsilon f_0 }
\right)S_2^2\right) = 1 + \alpha\left(
4V-\dph^2+\rhm-3\prm- 12\left( {b + \varepsilon f_0 }
\right)S_2^2\right)$, $\alpha \equiv f/({3f_0^2 })$, $\veps  \equiv q_2/f$
(hence, $q_2 = 3\alpha\, \veps f_0^2$). By using (\ref{13})--(\ref{16}) and the following relation
obtained from
 definition of $A_3$ and $A_4$
\begin{equation}\label{19}
A_{3}  + A_{4}  = \dot S_2  + 3HS_2  - 4S_1 S_2,
\end{equation}
we find for the torsion function $S_1$ from (\ref{10}) the following
expression:
\begin{equation} \label{20}
    S_1  = -\frac{3\alpha }{4Z} \left[
            \frac{\pat V}{\pat \phi} \dot\phi \ +H\left({Y + 2\dph^2}\right)
            -4\left( {2b - \veps f_0 } \right) S_2 \,\dot S_2
        \right],
\end{equation}
where
\[
    Y \equiv \left(\rhm+\prm\right)%
\left(3\frac{d\prm}{d\rhm}-1 \right) + 12\varepsilon
        f_0 S_2^2.
\]
Then by using formulas (\ref{16}) and (\ref{19}) we find from (\ref{11}) the following second order
differential equation for the torsion function $S_2$:
\begin{eqnarray}\label{21}\fl
    \varepsilon \left[ \ddot S_2  + 3H\dot S_2  + 3\dot{H}S_2  - 4\left(\dot S_1  - 3 HS_1
        + 4S_1^2\right) S_2  \right]
\nonumber \\
        - \frac{1} {{3f_0 }}\left( {\rho  - 3p - 12bS_2^2 } \right)S_2
        - \frac{{\left( {f_0  - b}\right)}} {f}S_2  = 0\,.
\end{eqnarray}
The obtained expressions (\ref{17})--(\ref{18}) for curvature functions $A_2$ and $A_1$ together
with their definition (\ref{2}) give the generalization of cosmological Friedmann equations for
HIM:
\begin{eqnarray}\label{22}\fl
    \frac{k}{R^2} + (H-2S_1)^2= \frac{1} {{6f_0 Z}}
        \left[
            {\rho  +6\left(f_0 Z- b\right) S_2^2
            + \frac{\alpha }{4} \left( {\rho  - 3p - 12bS_2^2 } \right)^2 }
        \right]
\nonumber\\
        - \frac{{3\alpha \varepsilon f_0 }} {Z}
            \left[
                {\left( {HS_2  + \dot S_2 } \right)^2
                + 4\left( {\frac{k}{{R^2 }} - S_2^2 } \right)S_2^2 }
            \right],
\end{eqnarray}
\begin{eqnarray}\label{23}\fl
    \dot{H}+H^2-2HS_1-2\dot{S}_1 = -\frac{1} {{12f_0 Z}}
        \left[
            \rho  + 3p - \frac{\alpha } {2} \left( {\rho  - 3p - 12bS_2^2 } \right)^2
        \right]
\nonumber\\
        - \frac{\alpha \varepsilon }{Z}\left( {\rho  - 3p - 12bS_2^2 } \right)S_2^2
        + \frac{{3\alpha \varepsilon f_0 }} {Z}
            \left[ {\left( {HS_2  + \dot S_2 } \right)^2
                + 4\left( {\frac{k}{{R^2 }} - S_2^2 } \right)S_2^2 }
            \right].
\end{eqnarray}
 These equations contain the torsion function $S_1$ determined by (\ref{20}) and the torsion
function $S_2$, satisfying the equation (\ref{21}). Obtained
equations contain three indefinite parameters: indefinite parameter
$\alpha$ determining the scale of extremely high energy densities
[4], parameter $b$ with dimension of parameter $f_0$ and the
parameter $\veps$ without dimension. We have to analyze all these
equations in order to investigate HIM with pseudoscalar torsion
function in the frame of PGTG.

\section{Asymptotics of cosmological solutions for HIM with pseudoscalar torsion function}

The structure of obtained equations (\ref{20})--(\ref{23}) describing HIM with two torsion
functions is essentially more complicated in comparison with the case of HIM with vanishing
function $S_2$. Note that if $S_2=0$ the equation (\ref{21}) vanishes and the cosmological
equations (\ref{22})--(\ref{23}) are transformed to GCFE containing the only indefinite parameter
$\alpha$ [4,5].

Now we will analyze the following question: by what restrictions on indefinite parameters the
cosmological solutions for HIM with pseudoscalar torsion function have the asymptotics in agreement
with actual observations. By taking into account that various parameters of HIM have to be small at
asymptotics, when values of energy density are sufficiently small, we see from (\ref{21}), that if
$|\veps|\ll 1$, the pseudoscalar torsion function has at asymptotics the following value:
\begin{equation}\label{24}
S_2^2  = \frac{{f_0(f_0  - b)}} {{4fb}} + \frac{{\rho  - 3p}}
{{12b }}.
\end{equation}
 Then we have at asymptotics: $Z \to (b/f_0)$, $S_1\to  0$ and the cosmological equations  (\ref{22})--(\ref{23}) at
 asymptotics take the form of cosmological Friedmann equations with cosmological constant:
 \begin{equation}\label{25}
    \frac{k} {{R^2 }} + H^2  = \frac{1} {{6b }}\left[ {\rho  + \frac{{3\,\left( {f_0  - b} \right)^2}}
         {{4f}}} \right],
\end{equation}
\begin{equation}\label{26}
    \dot H + H^2  =  - \frac{1} {{12b }}\left[ {\rho  + 3p - \frac{{3\left( {f_0  - b} \right)^2 }}
        {{2f}}} \right].
\end{equation}
From equations (\ref{25})--(\ref{26}) we see, that parameter $b$ has to be very close to $f_0$, but
smaller than  $f_0$. The value of $b$ leading to observable acceleration of cosmological expansion
depends on  the scale of extremely high energy density defined by $\alpha^{-1}$. If we take into
account  that the value of $\frac{3}{4}(f_0-b)^2/f= \frac{1}{4}\alpha^{-1}(1-b/f_0)^2$ is equal
approximately  to $0{.}7\rho_{\mathrm{cr}}$  (the critical energy density is
$\rho_{\mathrm{cr}}=6f_0 H_0^2$, where $H_0$ is the value of the Hubble parameter at present
epoch), then we obtain that $b=[1-(2{.}8 \rho_{cr}\alpha)^{1/2}]f_0$. If we suppose that the scale
of extremely high energy densities is larger than the energy density for quark-gluon matter, but
less than the Planckian energy density, then we obtain the corresponding estimation for $b$, which
is very close to $f_0$. Obtained restrictions on indefinite parameters will be used below for
investigation of inflationary cosmological solutions.

\section{Regular inflationary cosmological solutions with two torsion functions}

To obtain cosmological solution by integrating cosmological equations we have to use the equation
of state of gravitating matter, which is different at different stages of cosmological evolution.
So, at asymptotics one uses usually equation of state for dust matter ($\rho = 0$). At the same
time, in order to obtain cosmological solution for inflationary HIM, we will use at the beginning
of cosmological expansion the expressions (\ref{13}) for energy density and pressure by including
scalar field as one component of gravitating matter. Like GR, the inflationary stage appears, if
the value of scalar fields at the beginning of cosmological expansion is sufficiently large
($\phi>1\,M_{\mathrm{p}}$, $M_{\mathrm{p}}$ is Planckian mass) [10].

In order to investigate inflationary cosmological solutions at extremely high energy densities, by
using (\ref{13})--(\ref{15}) and (\ref{20}) we transform cosmological equations
(\ref{22})--(\ref{23}) and equation (\ref{21}) for $S_2$-function to the following form
\begin{eqnarray}\label{27}\fl
    H^2\left\{
            \left[ Z+\frac{3}{2}\alpha\left( Y+2\dph \right) \right]^2
            +3\alpha\veps f_0 S_2^2 Z
        \right\}
\nonumber\\
    +6\alpha H\left\{
            \left[ Z+\frac{3}{2}\alpha\left( Y+2\dph^2 \right) \right]
            \times\left[
                \frac{\pat V}{\pat \phi}\dph
                -2\left( 2b-\veps f_0 \right) S_2\dot{S}_2
            \right]
            +\veps f_0 S_2 \dot{S}_2 Z
        \right\}
\nonumber\\
    +9\alpha^2\left[
            \frac{\pat V}{\pat \phi}\dph-2\left( 2b-\veps f_0 \right)S_2 \dot{S}_2
        \right]^2
    +3\alpha\veps f_0\left[
            \dot{S}_2^2+4\left( \frac{k}{R^2}-S_2^2 \right) S_2^2
        \right] Z
\nonumber\\
    -\frac{1}{6 f_0} \left[
            \rhm+\frac{1}{2}\dph+V-6b S_2^2+\frac{1}{4}\alpha
                \left(
                    \rhm-3\prm+4V-\dph^2-12b S_2^2
                \right)^2
        \right] Z
\nonumber\\
    +\left( \frac{k}{R^2}-S_2^2 \right) Z^2=0,
\end{eqnarray}
\begin{eqnarray}\label{28}\fl
    \dot{H}\left[
            1 + \frac{3\alpha}{2Z}\left( Y + 2\dph^2 \right)
        \right]
    +H^2\left\{
            1 + \frac{3\alpha}{2Z}\left( Y + 2\dph^2 \right)
            -\frac{9\alpha^2}{2Z^2}\left( Y + 2\dph^2 \right)
                \left( Y + 2\dph^2 - 12 \veps f_0 S_2^2 \right)
\right.\nonumber\\
\left.
            -\frac{9\alpha}{2 Z}\left[
                    3\frac{d^2 \prm}{d\rhm^2} \left( \rhm + \prm \right)^2
                    +\left( 3 \frac{d\prm}{d\rhm}-1 \right)
                        \left( 1 + \frac{d\prm}{d\rhm} \right) \left( \rhm + \prm \right)
                    +4\dph^2
                \right]
        \right\}
\nonumber\\
    -\frac{3\alpha}{Z} H\left\{
            \left[4 \frac{\pat V}{\pat \phi}\dph +  2\left( 2b - 7\veps f_0 \right)S_2\dot{S}_2 \right]
            +\frac{3\alpha}{Z}\left[
                    \left(
                        \frac{\pat V}{\pat\phi}\dph - 2\left( 2b - \veps f_0 \right)S_2\dot{S}_2
                    \right)
\right.\right.\nonumber\\
\left.\left.\times
                    \left(Y+2\dph^2 - 12\veps f_0 S_2 \dot{S}_2\right)
                    +\left( Y + 2\dph^2 \right)
                            \times\left(
                                \frac{\pat V}{\pat\phi}\dph
                                -4\left( b + \veps f_0 \right)S_2\dot{S}_2
                            \right)
                \right]\right\}
\nonumber\\
    +\frac{3\alpha}{Z}\left\{
            \frac{\pat^2 V}{\pat\phi^2}\dph^2-\left( \frac{\pat V}{\pat\phi} \right)^2
\right.\nonumber\\
\left.
            -\frac{6\alpha}{Z}\left(
                    \frac{\pat V}{\pat\phi}\dph - 2\left( 2b-\veps f_0 \right) S_2\dot{S}_2
                \right)
                \times\left(
                    \frac{\pat V}{\pat\phi}\dph - 4\left( b+\veps f_0 \right) S_2\dot{S}_2
                \right)
\right.\nonumber\\
\left.
            -2\left( 2b - \veps f_0 \right) \left(\dot{S}_2^2 + S_2\ddot{S}_2 \right)
        \right\}
\nonumber\\
    =-\frac{1}{12f_0 Z}\left[
            \rhm + 3\prm -2\left( V - \dph^2 \right)
            - \frac{1}{2}\alpha \left( \rhm - 3\prm + 4V - \dph^2 - 12b S_2^2 \right)^2
        \right]
\nonumber\\
    -\frac{\alpha\veps}{Z}\left( \rhm - 3\prm + 4V - \dph^2 - 12b S_2^2 \right) S_2^2
\nonumber\\
    +3\frac{\alpha\veps f_0}{Z}\left[
            \left( H S_2^2 + \dot{S}^2 \right)^2 + 4\left( \frac{k}{R^2} - S_2^2 \right) S_2^2
        \right],
\end{eqnarray}
\begin{eqnarray}\label{29}\fl
    \ddot{S}_2\left[1 - \frac{12\alpha}{Z}\left( 2b - \veps f_0 \right) S_2^2 \right]
    +3\dot{H} S_2\left[1 + \frac{\alpha}{Z}\left( Y + 2\dph^2 \right) \right]
\nonumber\\
    -9\frac{\alpha}{Z} H^2S_2\left[
            Y + 6\dph^2 + 3\frac{d^2\prm}{d\rhm^2}\left( \rhm + \prm \right)^2
\right.\nonumber\\
\left.
            +\left( 3\frac{d\prm}{d\rhm} - 1 \right)\left( 1 + \frac{d\prm}{d\rhm} \right)
                \left( \rhm + \prm \right)
            +\frac{\alpha}{Z}\left( Y + 2\dph^2 \right)
\right.\nonumber\\
\left.\vphantom{\frac{d\prm}{d\rhm}}
\times
                \left( Y + 2\dph^2 - 12\veps f_0 S_2^2\right)
        \right]
    +3H S_2\left\{
        1 - 4\frac{\alpha}{Z} \left(
                4\frac{\pat V}{\pat\phi}\dph - 3\left( 2b + \veps f_0 \right) S_2\dot{S}_2
            \right)
\right.\nonumber\\
\left.
    -6\frac{\alpha^2}{Z^2} \left[
            \left(
                \frac{\pat V}{\pat\phi}\dph - 2\left( 2b - \veps f_0 \right) S_2\dot{S}_2
            \right)
            \left( Y + 2\dph^2 - 12\veps f_0 S_2 \dot{S}_2\right)
\right.\right.\nonumber\\
\left.\left.
            +\left( Y + 2\dph^2 \right)\left(
                    \frac{\pat V}{\pat\phi}\dph - 4\left( b + \veps f_0 \right) S_2\dot{S}_2
                \right)
            \right]\right\}
\nonumber\\
    -9\frac{\alpha^2}{Z^2} S_2\left[
            H\left(Y + 2\dph^2 \right)
            +2\left(
                    \frac{\pat V}{\pat\phi}\dph - 2\left( 2b - \veps f_0 \right) S_2\dot{S}_2
                \right)
        \right]^2
\nonumber\\
    -6\frac{\alpha}{Z} S_2\left[
            \left( \frac{\pat V}{\pat\phi} \right)^2 - \frac{\pat^2 V}{\pat\phi^2}\dph^2
            +2\left( 2b - \veps f_0 \right) \dot{S}_2^2
\right.\nonumber\\
\left.
            +\frac{6\alpha}{Z}\left(
                    \frac{\pat V}{\pat\phi}\dph - 2\left( 2b - \veps f_0 \right) S_2\dot{S}_2
                \right)
                \left(
                    \frac{\pat V}{\pat\phi}\dph - 4\left( b + \veps f_0 \right) S_2\dot{S}_2
                \right)
        \right]
\nonumber\\
    -\frac{1}{\veps}\left[
            \frac{1}{3f_0}\left( \rhm - 3\prm + 4V - \dph^2 - 12b S_2^2 \right)
            +\frac{f_0 - b}{f}
        \right] S_2
    =0.
\end{eqnarray}
Equation (\ref{27}) can be written as
\begin{equation}\label{30}
    AH^2 + 2BH + C = 0,
\end{equation}
where
\begin{eqnarray}\fl
    A = \left[ Z + \frac{3}{2}\alpha\left( Y + 2\dph^2 \right) \right]^2
        + 3\alpha\veps f_0 S_2^2 Z,
\nonumber
\end{eqnarray}
\begin{eqnarray}\fl
    B = 3\alpha\left\{
            \left[ Z + \frac{3}{2}\alpha\left( Y + 2\dph^2 \right) \right]
            \left[
                \frac{\pat V}{\pat\phi}\dph - 2\left( 2b - 2\veps f_0 \right) S_2\dot{S}_2
            \right]
            +\veps f_0 S_2 \dot{S}_2 Z
        \right\},
\nonumber
\end{eqnarray}
\begin{eqnarray}\fl
    C = 9\alpha^2\left[
                \frac{\pat V}{\pat\phi}\dph - 2\left( 2b - \veps f_0 \right) S_2\dot{S}_2
            \right]^2
        +3\alpha\veps f_0 \left[
                \dot{S}_2^2 + 4\left( \frac{k}{R^2} - S_2^2 \right) S_2^2
            \right] Z
\nonumber\\
        -\frac{1}{6f_0}\left[
                \rhm + \frac{1}{2}\dph^2 + V - 6b S_2^2
                +\frac{1}{4}\alpha\left(
                        \rhm - 3\prm + 4V - \dph^2 - 12b S_2^2
                    \right)^2
            \right] Z
\nonumber\\
        +\left( \frac{k}{R^2} - S_2^2 \right) Z^2.
\nonumber
\end{eqnarray}
From (\ref{30}) we obtain two $H_{\pm}$-solutions for the Hubble parameter
\begin{equation}\label{31}
    H_{\pm}=\frac{-B \pm \sqrt{D}}{A},
\end{equation}
where $D = B^2 - 4AC$.

At asymptotics $H_-$-solutions and $H_+$-solutions describe the stages of cosmological compression
and expansion respectively [4]. The transition from $H_{-}$-solution to $H_{+}$-solution takes
place when $D=0$.

Now we will analyze extremum surfaces in space of independent
variables $\phi$, $\dot\phi$, $S_2$, $\dot{S}_2$, $\rhm$, in the
points of which the Hubble parameter vanishes $H=0$. Extremum
surfaces depend on indefinite parameters $\alpha$, $\veps$, $b$ and
in the case of open and closed models also on the scale factor $R$.
By denoting values of variables on extremum surfaces  by means of
index $0$, we obtain from (\ref{30}) the following equation for such
surfaces
\begin{eqnarray}\label{32}\fl
    \frac{1}{6f_0}\left[
            \rhmz + \frac{1}{2}\dph_0^2 + V_0 - 6b S_{20}^2
            +\frac{1}{4}\alpha\left(
                    \rhmz - 3\prmz + 4V_0 - \dph_0^2 - 12b S_{20}^2
                \right)^2
        \right] Z_0
\nonumber\\
    -9\alpha^2\left[
            \left( \frac{\pat V}{\pat\phi} \right)_{\!0}\dph_0
            - 2\left( 2b - \veps f_0 \right) S_{20}\dot{S}_{20}
        \right]^2
\nonumber\\
    -3\alpha\veps f_0 \left[
            \dot{S}_{20}^2 + 4\left( \frac{k}{R_0^2} - S_{20}^2 \right) S_{20}^2
        \right] Z_0
    -\left( \frac{k}{R_0^2} - S_{20}^2 \right) Z_0^2
    =0,
\end{eqnarray}
where
$$
    Z_0 = 1 + \alpha\left[
                \rhmz - 3\prmz + 4V_0 - \dph_0^2 - 12\left( b + \veps f_0 \right) S_{20}^2
            \right].
$$

The derivative of the Hubble parameter on extremum surfaces obtained from (\ref{28})--(\ref{29}) is
determined as
\begin{eqnarray}\label{33}\fl
    \dot{H}_0 Z_0^2\left\{
            1 + \alpha\left[
                    \rhmz - 3\prmz
                    +\frac{3}{2}\left( 3\left( \frac{d\prm}{d\rhm} \right)_{\!0} - 1 \right)
                        \left( \rhmz + \prmz \right)
                    +4V_0 + 2\dph_0^2
                \right]
        \right\}
\nonumber\\
    =\left[
            1+\alpha\left( \rhmz - 3\prmz + 4V_0 - \dph_0^2 - 36b S_{20}^2 \right)
        \right]
    \times\left\{
            \frac{Z_0}{2f_0}\left[
                    \frac{1}{2}\left( \rhmz - \prmz \right) + V_0
\right.\right.\nonumber\\
\left.\left.
                    - 4bS_{20}^2
                    + \frac{1}{4}\alpha \left(
                            \rhmz - 3\prmz + 4V_0 - \dph_0^2 - 12b S_{20}^2
                        \right)^2
                \right]
\right.\nonumber\\
\left.
            +3\alpha Z_0\left[
                    \left( \frac{\pat V}{\pat\phi} \right)_{\!0}^2
                    -\left( \frac{\pat^2 V}{\pat\phi^2} \right)_{\!0}\dph_0^2
                    +\left( 4b - 3\veps f_0 \right) \dot{S}_{20}^2
                    -4\veps f_0\left( \frac{k}{R_0^2} - S_{20}^2
                        \right) S_{20}^2
\right.\right.\nonumber\\
\left.\left.
                    -\frac{1}{3}\veps\left(
                            \rhmz - 3\prmz + 4V_0 - \dph_0^2 - 12b S_{20}^2
                        \right) S_{20}^2
                \right]
            -2\left( \frac{k}{R_0^2} - S_{20}^2 \right) Z_0^2
\right.\nonumber\\
\left.
            -108\alpha^2\veps f_0 S_{20} \dot{S}_{20}\left[
                    \left( \frac{\pat V}{\pat\phi} \right)_{\!0}\dph_0
                    -2\left( 2b - \veps f_0 \right) S_{20}\dot{S}_{20}
                \right]
        \right\}
\nonumber\\
    +6\alpha\left( 2b - \veps f_0 \right) S_{20}^2\left\{
            72\alpha^2\left[
                    \left( \frac{\pat V}{\pat\phi} \right)_{\!0}\dph_0
                    -2\left( 2b - \veps f_0 \right) S_{20}\dot{S}_{20}
                \right]
\right.\nonumber\\
\left.
            \times\left[
                    \left( \frac{\pat V}{\pat\phi} \right)_{\!0}\dph_0
                    -\left( 4b + \veps f_0 \right) S_{20}\dot{S}_{20}
                \right]
            +6\alpha Z_0\left[
                    \left( \frac{\pat V}{\pat\phi} \right)_{\!0}^2
                    -\left( \frac{\pat^2 V}{\pat\phi^2} \right)_{\!0}\dph_0^2
\right.\right.\nonumber\\
\left.\left.
                    +2\left( 2b - \veps f_0 \right) \dot{S}_{20}^2
                \right]
\right.\nonumber\\
\left.
            +\frac{1}{\veps} Z_0^2\left[
                    \frac{1}{3f_0}\left(
                            \rhmz - 3\prmz + 4V_0 - \dph_0^2 - 12b S_{20}^2
                        \right)
                    +\frac{f_0 - b}{f}
                \right]
        \right\}.
\end{eqnarray}
In the case of HIM without pseudoscalar torsion function $S_2$ the equation (\ref{32}) and the
formula (\ref{33}) simplify and take the form obtained in ref. [4]. As it was noted in [4,6], in
this case the most part of extremum surfaces play the role of bounce surfaces $(\dot{H}_0>0)$ for
scalar field potentials applying in theory of chaotic inflation. The different situation is in
considering case of extremum surfaces (\ref{32}). By given values of parameters $\alpha$ and
$\veps$\footnote{According to the conclusion obtained in Section~3 the parameter $b$ is very close
to $f_0$. As result we put below for numerical calculations $b=f_0$.} the bounce $(\dot{H}_0>0)$
takes place only in limited domain of extremum surfaces (\ref{32}) with negligibly small values of
$S_{20}$. In the case $S_{20}=0$ the equation of extremum surface (\ref{32}) and the expression
(\ref{33}) of derivative $\dot{H}_0$ are simplified and take the following form
\begin{eqnarray}\label{34}\fl
    \frac{1}{6f_0}\left[
            \rhmz + \frac{1}{2}\dph_0^2 + V_0
            +\frac{1}{4}\alpha\left( \rhmz - 3\prmz + 4V_0 - \dph_0^2 \right)^2
        \right] Z_0
\nonumber\\
    -9\alpha^2\left( \frac{\pat V}{\pat\phi} \right)_{\!0}^2 \dph_0^2
    -\frac{k}{R_0^2} Z_0^2
    =3\alpha\veps f_0 \dot{S}_{20}^2 Z_0,
\end{eqnarray}
\begin{eqnarray}\label{35}\fl
    \dot{H}_0 = \left\{
            \frac{1}{2f_0}\left[
                    \frac{1}{2}\left( \rhmz - \prmz \right) + V_0
                    +\frac{1}{4}\alpha\left( \rhmz - 3\prmz + 4V_0 - \dph_0^2 \right)^2
                \right]
\right.\nonumber\\
\left.
                +3\alpha\left[
                        \left( \frac{\pat V}{\pat\phi} \right)_{\!0}^2
                        -\left( \frac{\pat^2 V}{\pat\phi^2} \right)_{\!0}\dph_0^2
                        +\left( 4b - 3\veps f_0 \right) \dot{S}_{20}^2
                    \right]
                -\frac{2k}{R_0^2} Z_0
        \right\}
\\ 
    \times\left\{
            1 + \alpha\left[
                    \rhmz - 3\prmz
                    +\frac{3}{2}\left( 3\left( \frac{d\prm}{d\rhm} \right)_{\!0} - 1 \right)
                        \left( \rhmz + \prmz \right)
                    +4V_0 + 2\dph_0^2
                \right]
        \right\}^{-1}.
        \nonumber
\end{eqnarray}
We see from (\ref{35}) that the presence of $\dot{S}_{20}$ in this expression does not prevent from
the bounce realization. Moreover, if we put $\phi=0$ and $k=0$, from (\ref{34}) follows that
$\veps>0$.

\begin{figure}[htb!]
\begin{minipage}{0.48\textwidth}\centering{
 \includegraphics[width=\linewidth]{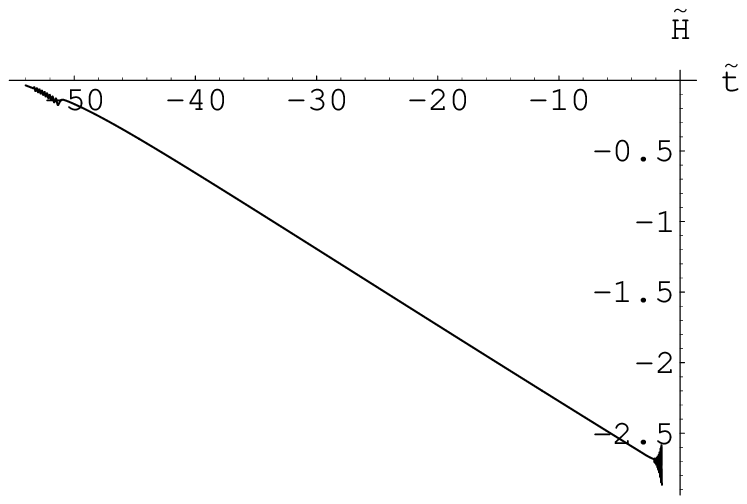}
}
\end{minipage}\, \hfill\,
\begin{minipage}{0.48\textwidth}\centering{
 \includegraphics[width=\linewidth]{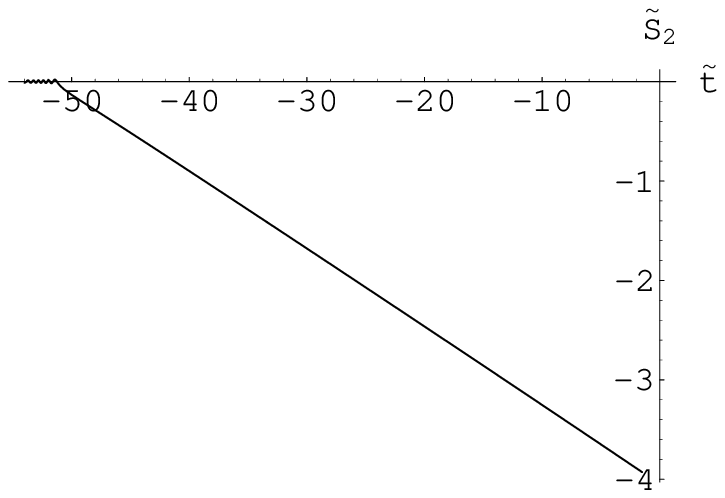}
}
\end{minipage}\\
%
\begin{minipage}{0.48\textwidth}\centering{
 \includegraphics[width=\linewidth]{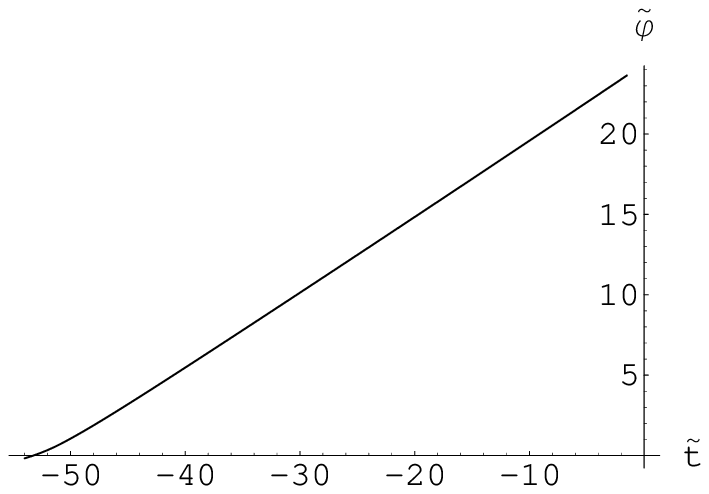}
}
\end{minipage}\, \hfill\,
\begin{minipage}{0.48\textwidth}\centering{
 \includegraphics[width=\linewidth]{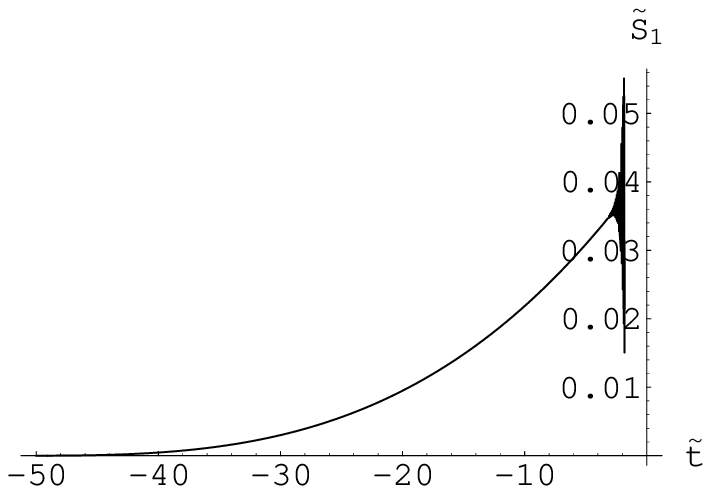}
}
\end{minipage}
\caption{\label{figm1}Compression stage.}
\end{figure}

\begin{figure}[htb!]
\begin{minipage}{0.48\textwidth}\centering{
 \includegraphics[width=\linewidth]{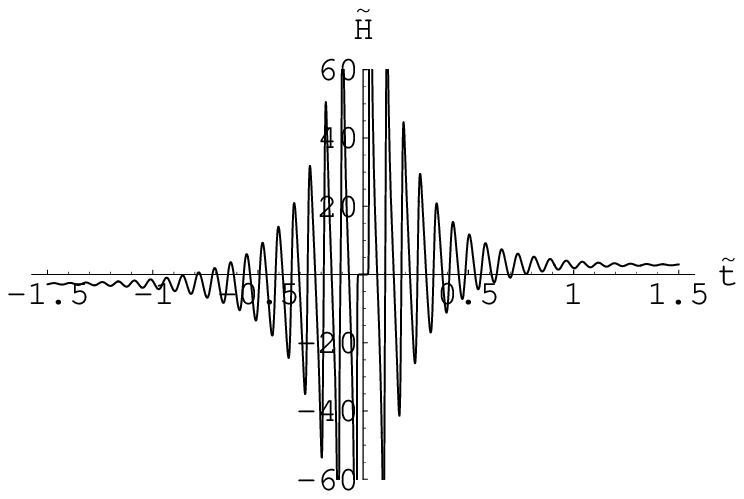}
}
\end{minipage}\, \hfill\,
\begin{minipage}{0.48\textwidth}\centering{
 \includegraphics[width=\linewidth]{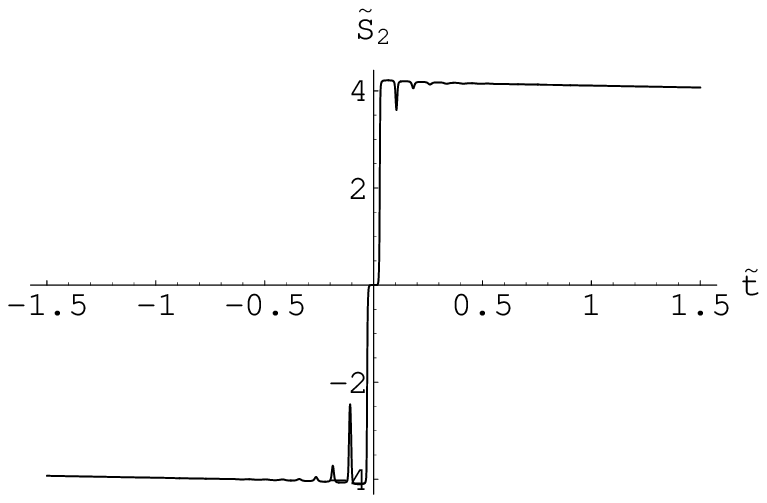}
}
\end{minipage}\\
%
\begin{minipage}{0.48\textwidth}\centering{
 \includegraphics[width=\linewidth]{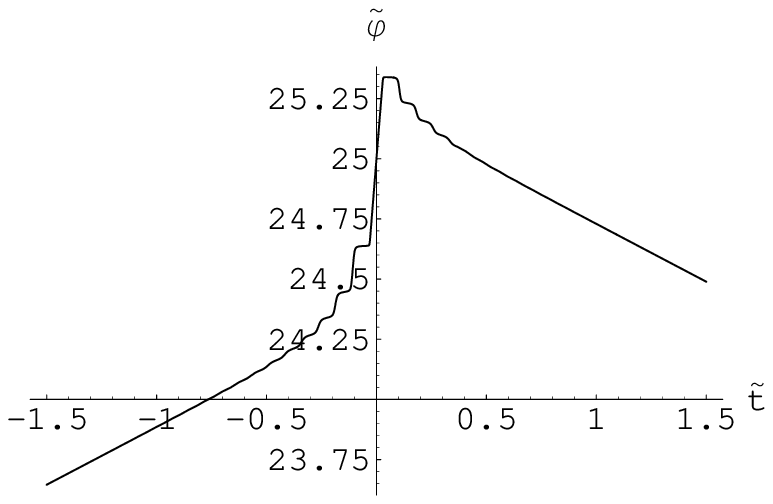}
}
\end{minipage}\, \hfill\,
\begin{minipage}{0.48\textwidth}\centering{
 \includegraphics[width=\linewidth]{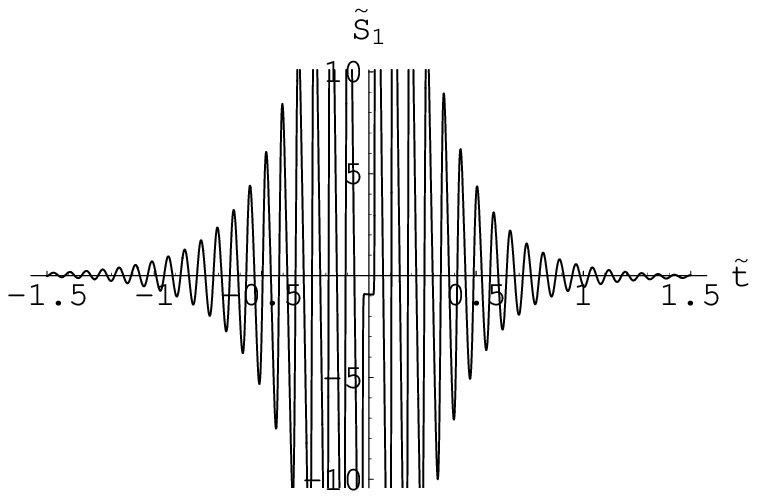}
}
\end{minipage}
\caption{\label{figm2}Transition stage.}
\end{figure}

\begin{figure}[hbt!]
\begin{minipage}{0.48\textwidth}\centering{
 \includegraphics[width=\linewidth]{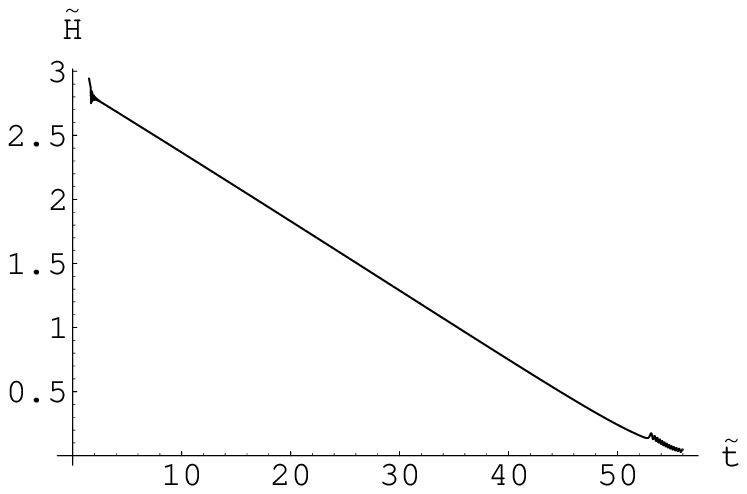}
}
\end{minipage}\, \hfill\,
\begin{minipage}{0.48\textwidth}\centering{
 \includegraphics[width=\linewidth]{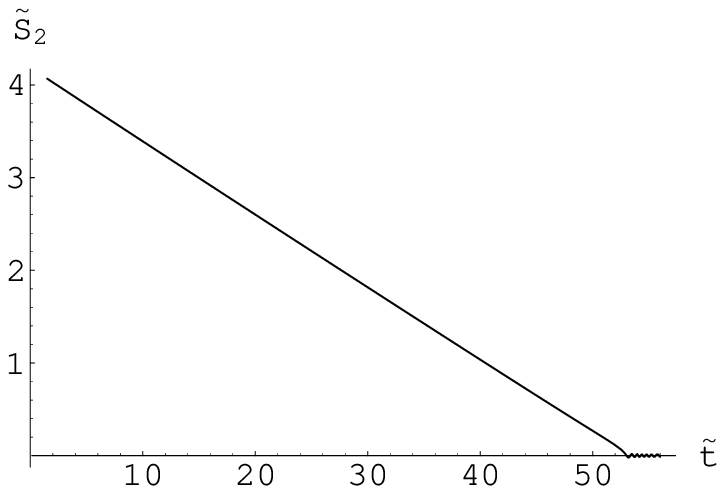}
}
\end{minipage}\\
%
\begin{minipage}{0.48\textwidth}\centering{
 \includegraphics[width=\linewidth]{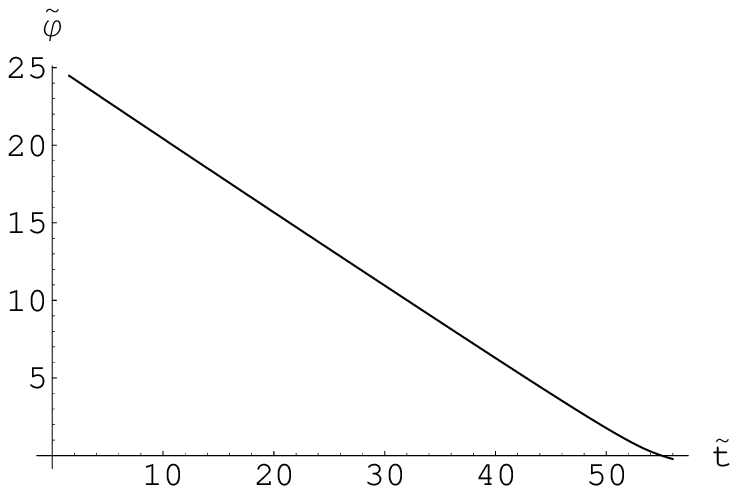}
}
\end{minipage}\, \hfill\,
\begin{minipage}{0.48\textwidth}\centering{
 \includegraphics[width=\linewidth]{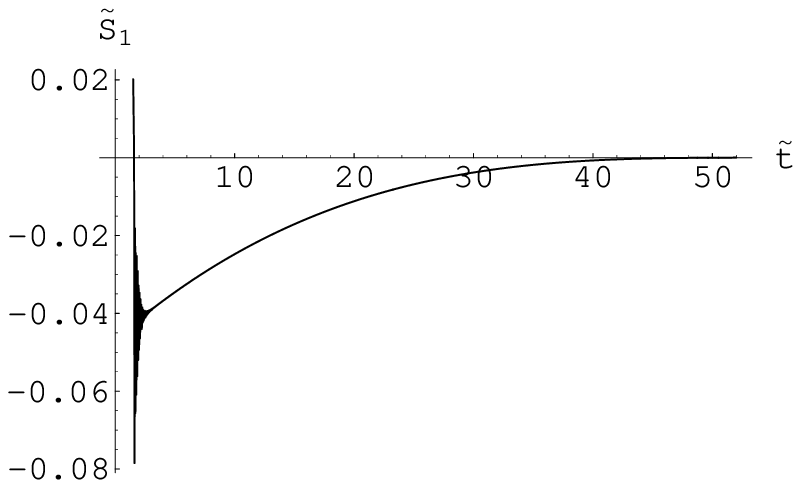}
}
\end{minipage}
\caption{\label{figm3}Inflationary stage.}
\end{figure}

\begin{figure}[hbt!]
\begin{minipage}{0.48\textwidth}\centering{
 \includegraphics[width=\linewidth]{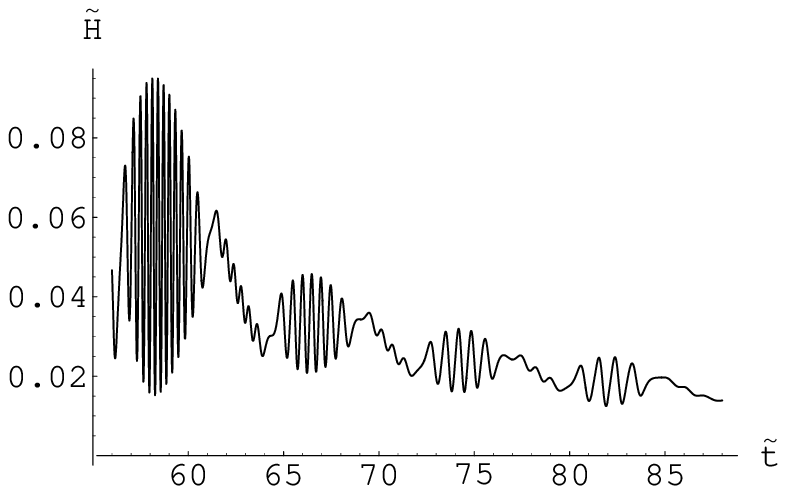}
}
\end{minipage}\, \hfill\,
\begin{minipage}{0.48\textwidth}\centering{
 \includegraphics[width=\linewidth]{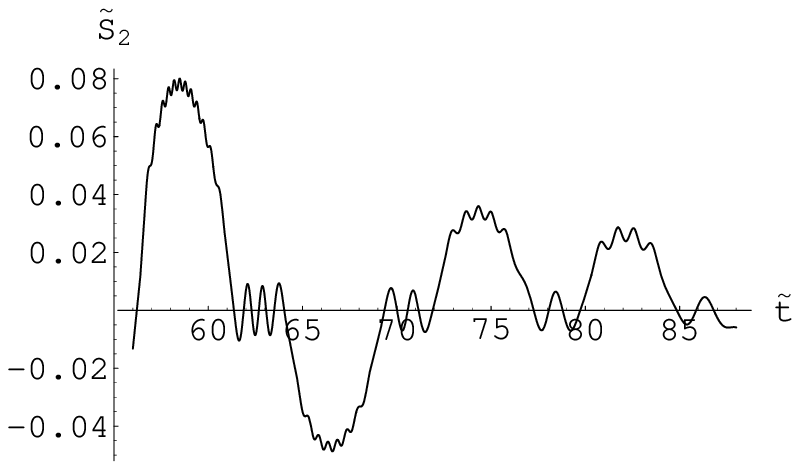}
}
\end{minipage}\\
%
\begin{minipage}{0.48\textwidth}\centering{
 \includegraphics[width=\linewidth]{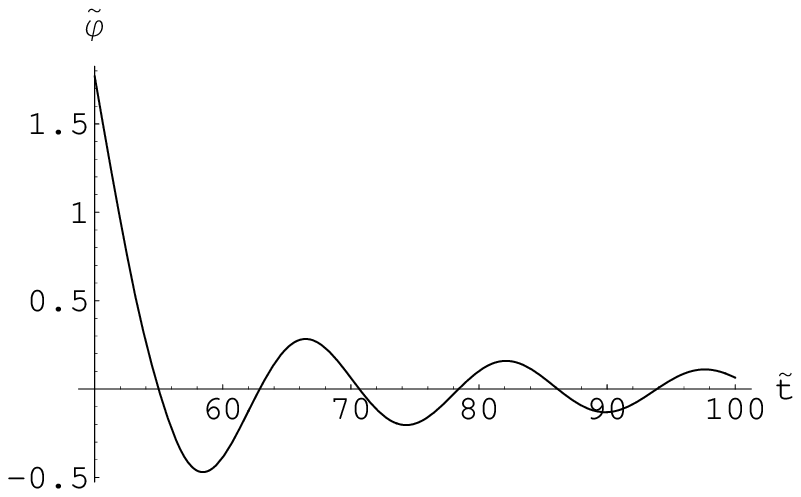}
}
\end{minipage}\, \hfill\,
\begin{minipage}{0.48\textwidth}\centering{
 \includegraphics[width=\linewidth]{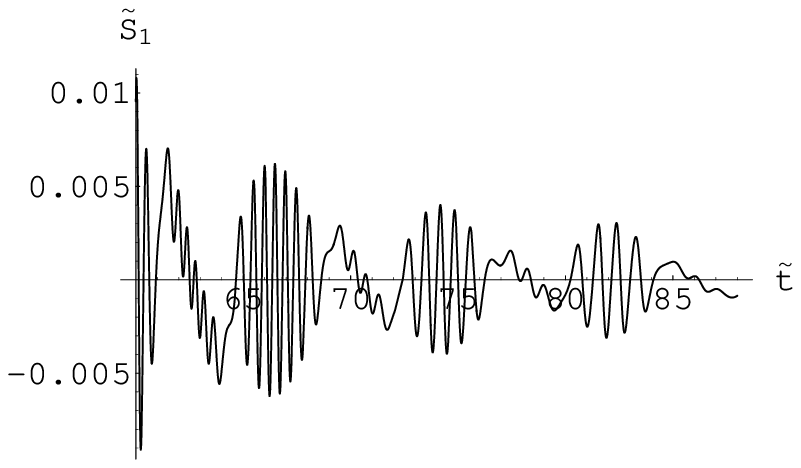}
}
\end{minipage}
\caption{\label{figm4}Postinflationary stage.}
\end{figure}

As an example of inflationary cosmological solutions we will consider below flat HIM filled with
ultrarelativistic matter $\prm=\frac{1}{3}\rhm$ and scalar field with quadratic potential
$V=\frac{1}{2}m^2\phi^2$.  For numerical calculations we will use $m=10^{-6}M_{\mathrm{P}}$ and
$\alpha^{-1}=1{.}2\times 10^{-13} M_{\mathrm{P}}^4$. To perform the numerical integration of
equations (\ref{14})--(\ref{15}), (\ref{28})--(\ref{29}) it is convenient to transform all
variables and parameters entering these equations to dimensionless units marked by tilde
\begin{equation}\label{dimless}
\begin{array}{lcl}
    t\to\tilde{t}=t/\sqrt{f_0 \alpha},& \qquad\qquad
            & R\to\tilde{R}=R/\sqrt{f_0 \alpha},\\
    \rho\to\tilde{\rho}=\alpha\,\rho, & & p\to\tilde{p}=\alpha\,p,\\
    \phi\to \tilde{\phi} = \phi/\sqrt{f_0}, & &
            m\to \tilde{m} = m\sqrt{f_0 \alpha}, \\
    H\to\tilde{H}=H\sqrt{f_0 \alpha}, & &
            S_{1,2}\to\tilde{S}_{1,2}=S_{1,2}\sqrt{f_0 \alpha}.
\end{array}
\end{equation}
The explicit form of equations (\ref{14})--(\ref{15}), (\ref{28})--(\ref{29}) after this
transformation is similar to original form except the fact that parameters $\alpha$ and $f_0$ are
cancelled in obtained equations. Particular numerical solution  was found under the following value
of indefinite parameter $\veps=10^{-4}$. Initial conditions for $\tilde{H}$, $\tilde{\phi}$,
$\tilde{S}_2$, $\tilde{S}'_2$, $\tilde{\rho}_{\mathrm{m}}$ were taken at a bounce as follows
\[
    \tilde{H}_0=0,\quad \tilde{\phi}_0=25, \quad \tilde{S}_{20}=0, \quad \tilde{S}'_{20}=0{.}001,
    \quad \tilde{\rhm}_{0}=0{.}4,
\]
where the prime denotes the differentiation with respect to $\tilde{t}$. Initial condition for
$\tilde{\phi}'_0$ was taken to satisfy (\ref{34}). Obtained solution is given in figure~1 --
figure~\ref{figm4} and includes four stages: the compression stage (figure~\ref{figm1}), the
transition stage from compression to expansion (figure~\ref{figm2}), the inflationary stage
(figure~\ref{figm3}) and the postinflationary stage (figure~\ref{figm4}). The distinguishing
features of obtained solution are its completely regular character. Note, that during inflationary
stage number of e-folds for the scale factor is equal approximately to $76$.

Similar to GR, the transition to radiation dominated stage can be realized by transformation of
oscillating scalar fields (see figure~4) into particles [11]. Details of such transition in
considered theory require further investigation. In particular, the presence of the oscillations of
the Hubble parameter $H$ (figure~4) can lead to some distinguishing features of the inflationary
scenario in the considered theory in comparison with the inflationary scenario in GR.

After transition to matter dominated stage the further evolution of the Universe in this theory is
the same as in the frame of standard cosmological scenario. The transition to the accelerating
expansion takes place, when the value of effective cosmological constant is greater than the matter
energy density.

\section{Conclusion}

As it was shown, in the framework of PGTG the gravitational interaction in the case of usual
gravitating systems can have the repulsion character not only at extreme conditions [4,5], but also
at sufficiently small energy densities. The pseudoscalar torsion function in HIM provokes the
appearance of effective cosmological constant at asymptotics of cosmological solutions that can
lead to observable accelerating cosmological expansion. Quantitative agreement of the obtained
result with observations depends on corresponding restrictions on indefinite parameters $\alpha$,
$b$ and $\veps$ from Section~3. Numerical solution for inflationary cosmological model presented at
Figures~1--4 conserves its qualitative behaviour by relatively small variations of indefinite
parameters and initial conditions.

The effect of acceleration of cosmological expansion in PGTG has the geometrical nature and is
connected with geometrical structure of physical space-time. Hence, from the point of view of
considered theory hypothetical form of gravitating matter
--- dark energy --- is fiction.

\end{document}